\documentclass[sigconf,nonacm, screen]{acmart}

\usepackage{longtable}
\usepackage{graphicx}
\usepackage[ruled]{algorithm}
\usepackage{algorithmic}
\usepackage{amssymb}
\usepackage{amsmath}
\usepackage{bm}
\usepackage{subcaption}
\newcommand{\argmax}{\mathop{\rm arg~max}\limits}

\AtBeginDocument{%
  \providecommand\BibTeX{{%
    \normalfont B\kern-0.5em{\scshape i\kern-0.25em b}\kern-0.8em\TeX}}}

\setcopyright{acmcopyright}
\copyrightyear{2020}
\acmYear{2020}
\acmDOI{}

\acmConference[SUM '20]{SUM '20: SUM}{Feb 07, 2020}{Houston, TX}
\acmBooktitle{SUM '20: State-based User Modelling, Feb 07, 2020, Houston, TX}
\acmPrice{}
\acmISBN{}

\begin{document}

\title{Fatigue-Aware Ad Creative Selection}


\author{Daisuke Moriwaki} 
\email{moriwaki\_daisuke@cyberagent.co.jp}
\author{Komei Fujita} 
\email{fujita\_komei@cyberagent.co.jp}
\author{Shota Yasui}
\email{yasui\_shota@cyberagent.co.jp}
\affiliation{%
\institution{CyberAgent, Inc.}}
    
\author{Takahiro Hoshino} 
\email{bayesian@jasmine.ocn.ne.jp}
\affiliation{%
\institution{Keio University \\RIKEN Center for Advanced Intelligence Project}}

\renewcommand{\shortauthors}{Moriwaki et al.}


\begin{abstract}
In online display advertising, selecting the most effective ad creative (ad image) for each impression is a crucial task for DSPs (Demand-Side Platforms) to fulfill their goals (click-through rate, number of conversions, revenue, and brand improvement). As widely recognized in the marketing literature, the effect of ad creative changes with the number of repetitive ad exposures. In this study, we propose an efficient and easy-to-implement ad creative selection algorithm that explicitly considers user's psychological status when selecting ad creatives. The proposed system was deployed in a real-world production environment and tested against the baseline algorithms. The results show superiority of the proposed algorithm.
\end{abstract}




\maketitle

\section{Introduction}
\label{sec:Introduction}
\begin{figure}[htb]
  \centering
  \includegraphics[width=\linewidth]{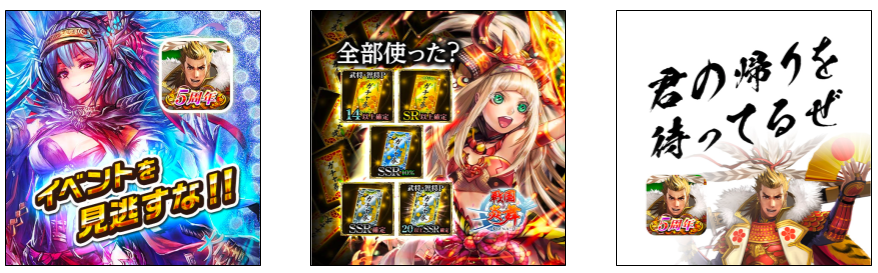}
  \caption{Variation of ad creatives for a mobile game title. The messages are all different:``Don't miss the special event!'' (left), ``Have you already tried these coupons?'' (center), and ``Waiting for your return'' (right). Users jump to login screen of the game when click on these ads.}
  \label{fig:creative}
  \Description{}
\end{figure}

In online display advertising, it is a common practice to prepare multiple ad creatives (a combination of an image and text or a video, see Figure \ref{fig:creative} for example) for each advertising campaign. The DSP (Demand-Side Platform) selects the most effective ad creative from the candidates.

Since new ad creatives are continuously added every day, every hour in the typical online advertising scene there is little time to test the effectiveness of each ad creative beforehand. The {\it bandit algorithm}\cite{lattimore_bandit_2019} is best suited to such situation. The bandit algorithms give us a way to choose one from multiple choices under uncertainty of the outcome by balancing exploration and exploitation.

The effectiveness of ad creative depends not only on the quality of ad creative but also on the audience. For example, one might prefer an informative ad creative while another like an emotional ad creative. The DSP, therefore, often considers {\it context} as well as the ad creative. The bandit algorithm that learn contexts is called {\it contextual bandit} and have been extensively studied at a practical level \cite{li_contextual-bandit_2010,chapelle_empirical_2011,tang_automatic_2013,bietti_contextual_2018}.

One serious side-effect of the contextual bandit is that the ad selection becomes deterministic once the algorithm has {\it learned} enough volume of data. The basic bandit algorithm, regardless of its implementation (e.g. UCB, Thompson sampling, or $\epsilon$-greedy), assumes the reward of arms are unchanged. The degree of exploration is decreasing with the number of plays and eventually goes to zero. As a result, targeted users keep watching same ad creative that is considered as the best by the algorithm given users' information.

\begin{figure}[htb]
  \centering
  \includegraphics[width=0.5\linewidth]{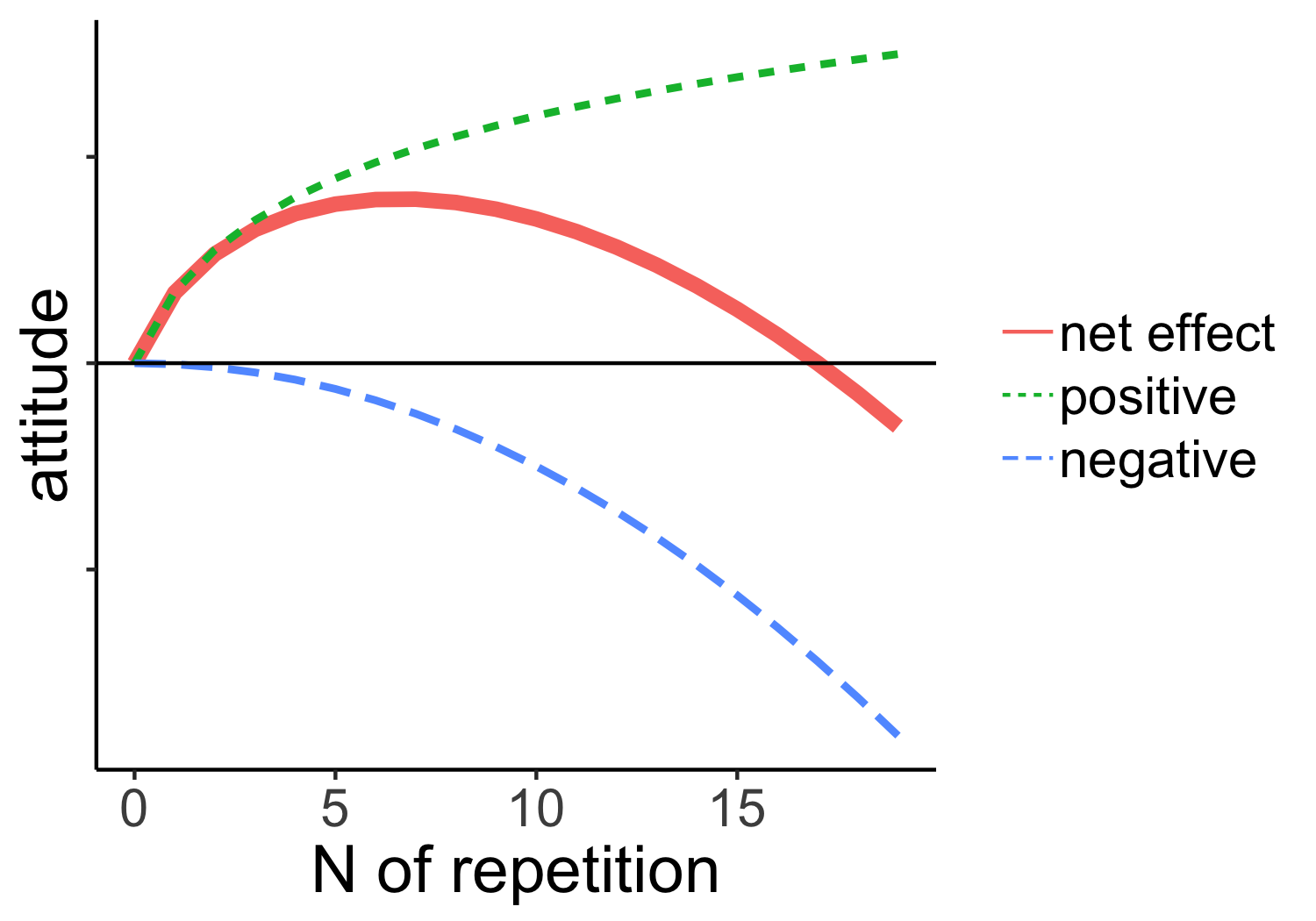}
  \caption{A graphical image of "Two-factor model" for advertising repetition. Previous researches have found that the repetitive advertising has both positive and negative effects on consumers' attitude toward advertised products and the net effect is described as an inverted U-shaped curve\cite{schmidt_advertising_2015}.}
  \label{fig:two factor}
  \Description{}
\end{figure}

The marketing science literature has been extensively studying the consequences of repetitive ad exposure. A seminal work by Zajonc\cite{zajonc_attitudinal_1968} shows mere repeated exposure can incur positive attitude among audience from the experiment results. On the other hand, many researchers point out there is a threshold above which additional ad exposure has negative effect for users' attitude due to boredom or advertising fatigue. As surveyed in \cite{pechmanAdvertisingRepetitionCritical1988,schmidt_advertising_2015}, {\it Two-factor model} theory is widely supported by the researchers in this domain. This theory claims that repetitive advertising has two phases: in the early stage the attitude toward the product improves with the number of repetition but at some point advertising fatigue dominates the positive effect and the net effect (positive effect minus negative effect) plummets (Figure \ref{fig:two factor}).

\begin{figure}[htb]
  \centering
  \includegraphics[width=0.5\linewidth]{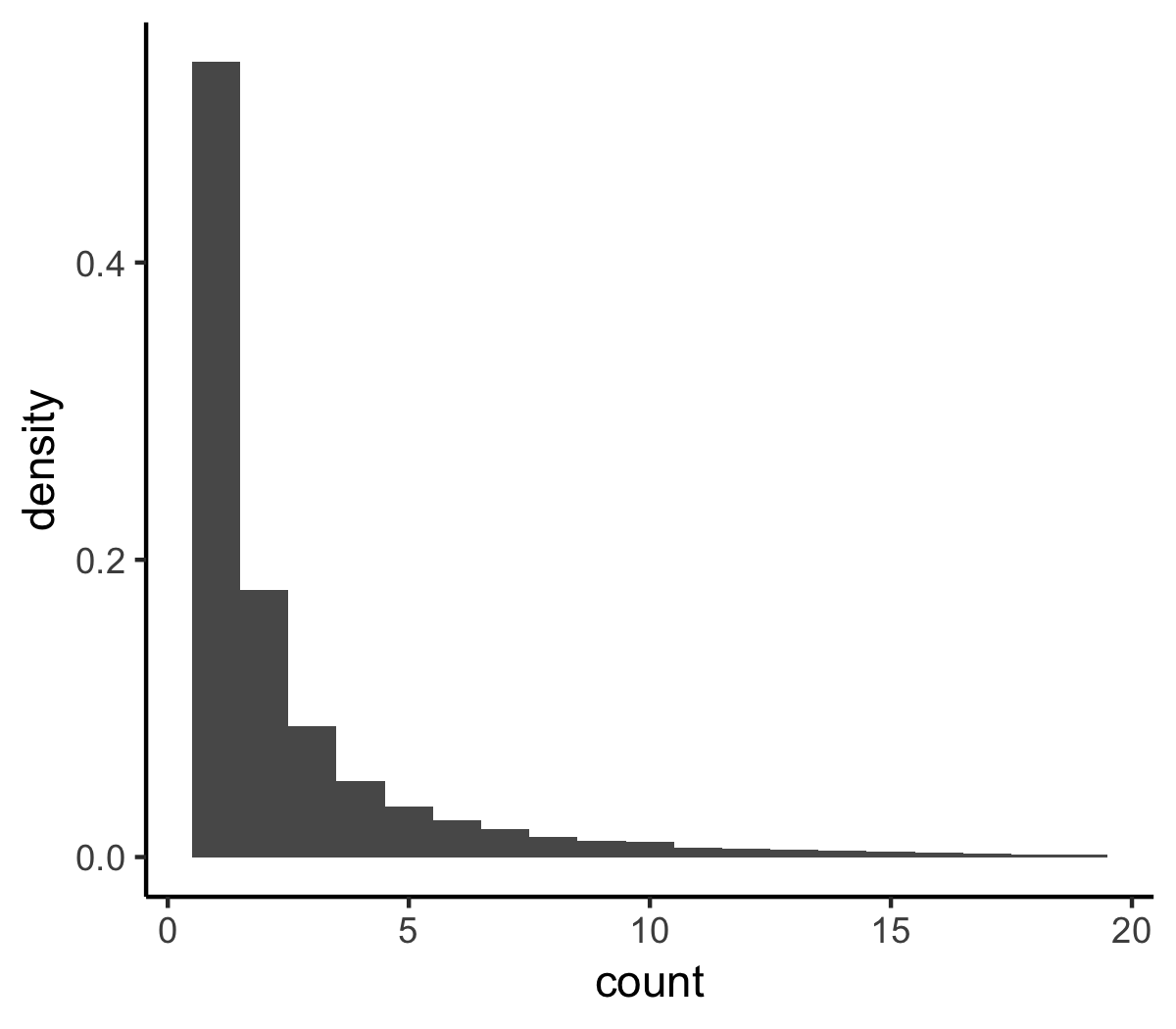}
  \caption{Number of exposure to same ad creative for each user within 24 hours. $N = 27,750$}
  \label{fig:adexposures}
  \Description{}
\end{figure}

The over-exposure of same ad is especially prevalent in display advertising. The real-time bidding (RTB) enables advertisers to reach out to targeted users whenever user browses web or use apps with advertising slots. The distribution of the counts of each user-ad creative pairs from real RTB data (Figure \ref{fig:adexposures}) shows that 47.5\% of creative-user pairs appear multiple times within 24 hours, which leads that 64.6\% of the users experience multiple exposure to same ad creative. 

\begin{figure}[htb]
  \centering
  \includegraphics[width=0.5\linewidth]{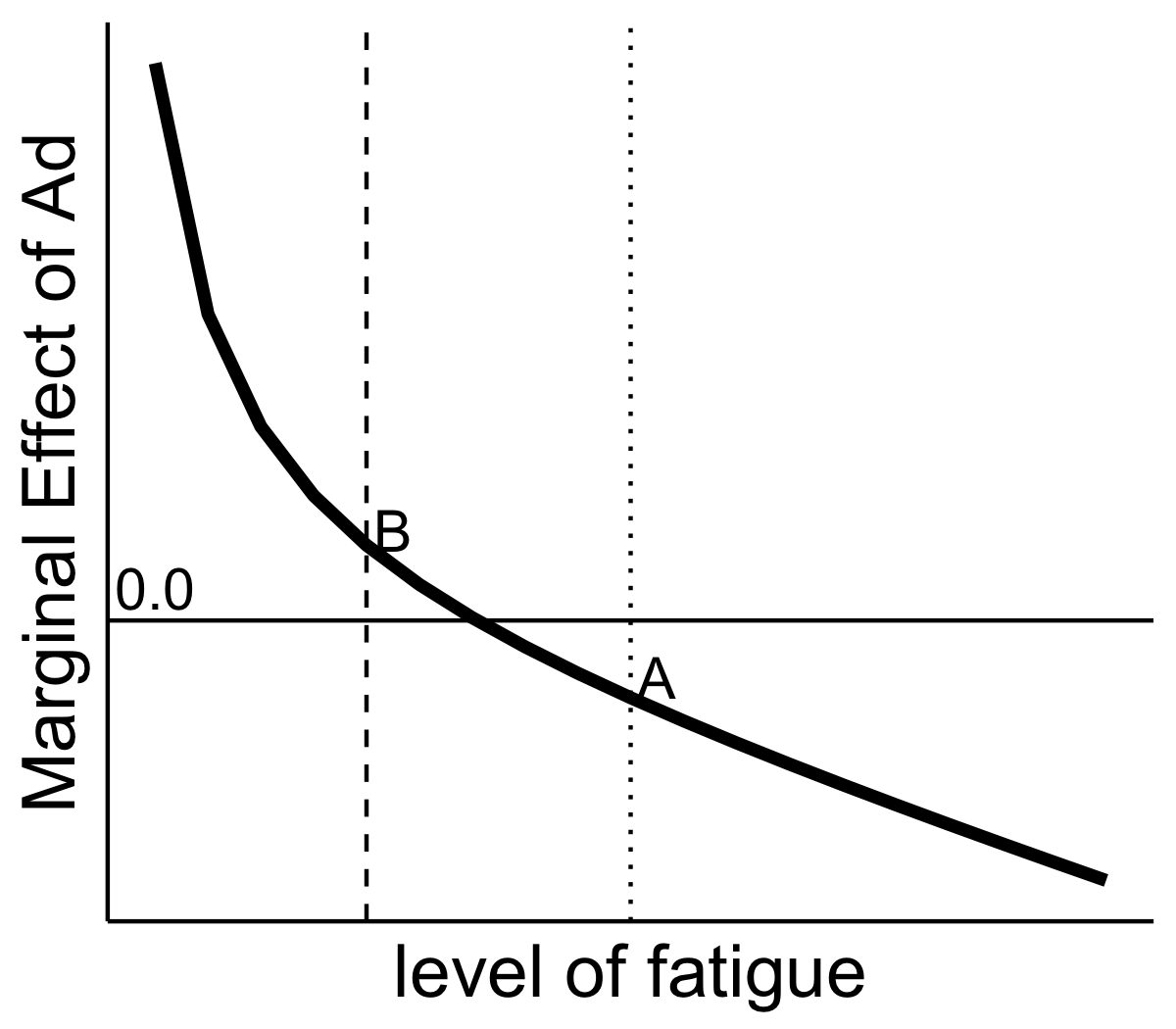}
  \caption{The level of fatigue and effect of additional ad exposure (a graphical image). According to the two-factor model, user who has seen same ad creative many times gets advertising fatigue and the effect of ad exposure is negative (A). But new ad creatives can reduce fatigue level and restore effectiveness (B).}
  \label{fig:marginaleffect}
  \Description{}
\end{figure}

Recently the concept of advertising fatigue has received growing attention from machine learning and data mining researchers\cite{maUserFatigueOnline2016a,guptaFactoringExposureDisplay2012,agarwalSpatiotemporalModelsEstimating2009a,abramsPersonalizedAdDelivery2007a}. In particular, Agarwal et al. (2009)\cite{agarwalSpatiotemporalModelsEstimating2009a} and Ma et al. (2016)\cite{maUserFatigueOnline2016a} proposed a fatigue-aware news recommendation system which evidently improved upon existing algorithm. Similar idea is applied to LinkedIn recommendation system\cite{leeModelingImpressionDiscounting2014}. The DSPs also need to consider the psychological status of the audience when choosing an ad creative. To capture users' perception toward each ad creative, we introduce a quantitative measure of advertising fatigue. When we have only one ad creative in hand, fatigue will be well proxied by the number of ad exposures in the past. However, it is not the case in display advertising. When we have multiple creatives, the past exposure to other creatives should be considered because the creatives more or less resemble to each other and expected to increase fatigue in a similar way. We conjecture the gain in fatigue is correlated with the degree of similarity of the ad creative on user's screen and what he has seen in the past. 
To get a more concrete idea, let's assume user's attitude toward ad creative evolves according to Figure \ref{fig:two factor}. Then the magnitude of the improvement of attitude by the additional ad exposure ($=$marginal effect of ad exposure) decreases as user's advertising fatigue increases (Figure \ref{fig:marginaleffect}). Suppose that a user has seen same ad creative repeatedly then the effect of the ad creative is negative due to the high level of fatigue (point A). In this case, we can restore the effect of advertising by using a new ad creative (point B). The new ad creative can be a slight modification of the original or totally different ones. A typical set of creatives is already shown in Figure \ref{fig:creative}. That is, the DSP needs to switch ad creative when user gets high level of fatigue.

We propose a new algorithm for ad creative selection for DSP that explicitly considers advertising fatigue. The algorithm not only selects the best ad creative based on the context but also estimate the level of fatigue for each user and take it into account.

Since the display advertising environment has strict latency requirements, the DSP server can spare only a few milliseconds for computation since it should deliver thousands of ad creatives to numerous users within unnoticeable duration. To save as much as computation time, we first develop a very simple calculation formula for advertising fatigue which capture the similarity of current ad creative and past ad exposure. Then we propose the implementable algorithm that add fatigue features to the existing reward predictor. To check the feasibility and the performance, we conducted an online experiment on real-world display advertising. The result shows that adding fatigue feature improves CTR (click-through rate, the percentage of being clicked out of impressions) and CVR (conversion rate, the percentage of users that complete desired goal such as log-in) over existing algorithm without any systemic problem. We also conducted post-experiment analysis and find the algorithm indeed reduced the fatigue. Therefore, our experiments show an apparent relationship between fatigue and CTR/CVR. 

The rest of the paper is as follows. In the next section, we review existing work in Section \ref{sec:Related Works}. Then we formalize the problem and explain the existing system in Section \ref{sec:Ad Creative Selection}. 
Section \ref{sec:Fatigue-Aware Ad Creative Selection} proposes the new algorithm that takes the advertising fatigue into account. 
Section \ref{sec:Evaluation} describes our online experiment and the results. Section \ref{sec:Discussion} conducts post-experiment analysis. Finally, Section \ref{sec:Conclusion} concludes.

\section{Related Works}
\label{sec:Related Works}
We review related works in both marketing science and machine learning.

\subsection{User fatigue in Machine Learning Literature}
Agarwal et al. (2009)\cite{agarwalSpatiotemporalModelsEstimating2009a} perform explanatory analysis on user's repeat-exposure to recommended articles using data from {\it Yahoo!} front page. They incorporate decaying factors to their CTR predictor based on the analysis. Ma et al. (2016)\cite{maUserFatigueOnline2016a} also conduct an extensive analysis for user's past exposure on data from {\it Bing} news recommendation engine. They propose to use users' past exposure and reaction to same news item and items from same category as features for the machine learning model. User fatigue is also considered explicitly in \cite{guptaFactoringExposureDisplay2012,abramsPersonalizedAdDelivery2007a}. The present paper extends their researches in a number of ways. First we extends the literature from recommendation to display advertising. Second, we define and calculate item similarity to measure fatigue. Third, we deployed our proposed algorithm in a production environment and proved its efficiency and feasibility. Not explicitly mentioning user fatigue, Lee et al. (2014)\cite{leeModelingImpressionDiscounting2014} improves recommendation algorithms in {\it LinkedIn} using user's past exposure to the recommended items (other members) and the actions to them.

\subsection{Repetitive Ad in Marketing Science}
The relationship between the number of repetitions and the effectiveness of ad have been actively studied in marketing science literature for a long time. Extensive reviews and meta-analysis point out that the optimal number of repeated ad exposures depends many conditions and thus is hard to determine\cite{pechmanAdvertisingRepetitionCritical1988,schmidt_advertising_2015}. In the context of online advertising, an analysis of large scale \textit{natural experiments} finds that  wear-out occurs in a heterogeneous manner\cite{lewis_worn-out_2015}. A lab experiment reveals that ad creatives with high quality is immune to wear-out effects\cite{chen_effects_2016}. 

\subsection{Effect of Varied Ad}
The effectiveness of varied ad creative has been examined in marketing science literature \cite{chatterjeeModelingClickstreamImplications2003,unnavaEffectsRepeatingVaried1991a,schumannPredictingEffectivenessDifferent1990}. In particular, Chatterjee et al. (2003)\cite{chatterjeeModelingClickstreamImplications2003} examined effectiveness of (i)repeating same ad creative and (ii)varying ad creative in lab experiments. The point estimate of CTR (click-through rate) of varied ad creative exceeded the same ad creative while their test statistics is not significant due to its small sample size.

\subsection{Bandit Algorithms in Online Ad/Recommendation System}
The bandit algorithms have been already applied to variety of advertising/recommendation system \cite{li_contextual-bandit_2010,chapelle_empirical_2011,tangEnsembleContextualBandits2014}. In particular, an application to ad format selection is closest to ours\cite{tang_automatic_2013}. Our research is also related to the {\it non-stationary bandit} problem, in which the reward is changing with time and the optimal arm is not always the same. Among various algorithms, directly estimating the decaying factor of rewards are proved to be effective\cite{liu_time-decaying_2014,levine_rotting_2017}. 
We also consider continuous creation and deletion of arms, which is considered in \cite{chakrabarti_mortal_nodate}. While existing literature focus on relatively simple mechanism of reward changing (e.g. decrease with time or pulls), we introduce a numerical measure of ad fatigue of users and tackle the problem. Several studies explicitly consider user abandoning due to the users' psychological status in contextual bandit\cite{lei_actor-critic_2017,cao_dynamic_2019}. In these studies, profitable interventions are considered to have a negative effect on users psychological status and possibly cause user churn. In this scenario, a good contextual bandit algorithm needs to save the frequency of uncomfortable treatments to keep the users engaged. The proposed algorithms solve a constrained optimization problem to find the best policy \cite{lei_actor-critic_2017} or calculate the optimal sequence of interventions \cite{cao_dynamic_2019} at each time of intervention. However, both studies are limited to the simulation and not applied to real production environment.

\section{Problem Setting and Existing Ad Creative Selection System}
\label{sec:Ad Creative Selection}

In this section we introduce real-world problem setting and explain current deployed system.

\subsection{Ad Creative Selection for Retargeting to Game Users}
In this paper, we focus on a DSP that targets the churned game users who have stopped playing for a while. 
But the same idea can be applied to a variety of settings in advertising and recommendation. 
The DSP intensively buys ad views of these users through Real-Time Bidding (RTB) and urge them to return to the game. The example of ad creatives has been already shown in Figure \ref{fig:creative}, which promote new features of the game, issuing coupons, or simply persuading them to return. Some of the users click on the ad, jump to the login screen, and login the game (conversion).

To convert users, the DSP needs to select the most effective ad creative for each user. The existing system first predicts CTR for each ad creative based on contextual information and then chooses one with the highest predicted CTR. 

Formally, each time $t$, the CTR $r_t$ is predicted based on the context vector $\mathbf{x}_t$ for ad creative $a_t$,
\begin{equation}
\hat{r_t}(a_t) = \sigma\left(\bm{\theta_0} \cdot \mathbf{x}_t + \bm{\theta}(a_t)\cdot \mathbf{x_t}\right),
\label{eq:basic}
\end{equation}
where $\bm{\theta_0}$ is a $d$-dimensional weight vector that is invariant to ad creative, and $\bm{\theta}(a_t)$ is a $d$-dimensional {\it action-specific} weight vector. Both vectors include bias term. $\sigma(\cdot)$ represents the sigmoid (logistic) function. The context vector $\mathbf{x}_t$ includes information from bid request (e.g. device, site/app and SSP), and hour of the day. The algorithm selects the ad creative with the highest expected reward.

One serious problem with the algorithm is that it does not generate good training data. The algorithm chooses same ad creative for same context. However, we need the results for the other ad creatives to train the model. The bandit algorithm is an established remedy to the problem that add randomization to selection algorithms. Following \cite{chapelle_empirical_2011}, the existing system adopts the Thompson sampling. That is, for each time $t$, action-specific weight vector $\bm{\theta}(a_t)$ is sampled from the distribution $\mathcal{N}(\bm{\mu}(a_t), \alpha \bm{\Sigma}(a_t))$ for each candidate ad creative, where scalar $\alpha \in (0,1]$ controls the degree of exploitation. Action-invariant vector $\bm{\theta}_0$ is set to $\bm{\mu}_0$ since it does not matter for selection.

Then the system chooses the best ad creative from candidates set $\mathcal{A}_t$, namely, $a_t^\ast = \argmax\limits_{a_t \in \mathcal{A}_t}\hat{r}_t(a_t)$. The action space $\mathcal{A}_t$ is different across $t$. $\mathcal{A}_t$ is determined based on the characteristics of user and ad-slot by the system. Moreover, the set of available ad creative changes over time. As explained in \cite{chakrabarti_mortal_nodate}, ads are continuously added and deleted from circulation. The algorithm chooses newly-added ad creative with a probability of $1/|A_t|$ for exploration. The whole procedure is described in Algorithm \ref{alg:contextual}.

\begin{algorithm}[htb]
\caption{Baseline Ad Creative Selection Algorithm (Baseline)}         
\label{alg:contextual}                          
\begin{algorithmic}                  
\REQUIRE $\mathcal{N}(\bm{\mu}(a), \alpha\bm{\Sigma}(a))$
\FOR{$t = 1, \cdots, T$}
\STATE Take feature vector $\mathbf{x}_{t}$, 
\FOR{$a_t = 1, \cdots, |\mathcal{A}_t|$}
\IF{$\bm{\mu}(a_t)$, $\bm{\Sigma}(a_t)$ is available}
\STATE Sample $\bm{\theta}(a_t) \sim \mathcal{N}(\bm{\mu}(a_t), \alpha\bm{\Sigma}(a_t))$
\STATE Calculate $\hat{r}_{t}(a_t)$ according to eq.(\ref{eq:basic}).
\ELSIF{$\bm{\mu}(a_t)$, $\bm{\Sigma}(a_t)$ is NOT available due to lack of data}
\STATE Assign $\hat{r}_t(a_t)$ very large value at the probability of $\frac{1}{|\mathcal{A}_t|}$ and very small values at the probability of $1-\frac{1}{|\mathcal{A}_t|}$
\ENDIF

\ENDFOR
\STATE Select $a^\ast = \argmax\limits_{a \in \mathcal{A}_t} \hat{r}_{t}(a)$
\STATE Observe $r_t \in \{0, 1\}$
\IF{$t$ is the last impression of the day}
\STATE Update $(\bm{\mu}(a), \bm{\Sigma}(a))$ using data from the previous day
\ENDIF
\ENDFOR
\end{algorithmic}
\end{algorithm}

\subsection{Parameter Estimation}
\label{sec:Parameter Estimation}
We estimate $\left(\bm{\mu}(a_t), \bm{\Sigma}(a_t)\right)$ as posterior Gaussian distribution of the weight vector $\bm{\theta(a_t)}$ of eq.\ref{eq:basic}. That is, we first estimate $\bm{\theta} = (\bm{\theta_0}, \bm{\theta}(a_t))$ as point estimates from L2-regularized logistic regression and derive $\bm{\Sigma_0}, \bm{\Sigma}(a_t) \in \bm{\Sigma}$ since the inverse of Hessian of the objective function (negative log-likelihood plus L2-penalty) becomes variance-covariance matrix of the weight vector.
Since the derivation is same as Chapell and Li (2011)\cite{chapelle_empirical_2011} we do not detail here. It is basically using the result that the L2-regularized logistic regression is equivalent to MAP estimation and utilize Laplace transformation (see Bishop (2006)\cite{bishopPatternRecognitionMachine2006}). Due to the intractable size of our Hessian matrix, we only use diagonal elements of Hessian.

To estimate both common weight parameters $\bm{\mu}_0$ and action-specific $\bm{\mu}(a)$, $a \in \mathcal{A}$ at once, we first expand feature vector $\mathbf{x}_t$ by interacting a binary action vector $\mathbf{a} \in \{0, 1\}^{|\mathcal{A}|}$ whose $k$th element takes one when corresponding action is taken while the others take zeros. That is, for $\mathbf{a}_t = (0,1,0)^T$, $\mathbf{a}_t \otimes \mathbf{x}_t = (\mathbf{0}_d^T, \mathbf{x}_t^T, \mathbf{0}_d^T)^T$ where $0_d$ denotes a zero vector of length $d$. Finally, by combining action-invariant component $\mathbf{x}_t$, we have $\mathbf{z_t} = (\mathbf{x}_t^T, \mathbf{a}_t \otimes \mathbf{x}_t)^T$. We use $\mathbf{z}_t$ for both training and predicting procedures. The actual implementation is detailed in Section \ref{sec:Evaluation}.

\section{Fatigue-Aware Ad Creative Selection}
\label{sec:Fatigue-Aware Ad Creative Selection}
As explained in Section \ref{sec:Introduction}, we need to consider user's psychological status when predicting reward. Formally, our {\it fatigue aware} reward predictor becomes
\begin{eqnarray}
\label{eq:predictor with fatigue} \hat{r}_{i, t}(a_t) = \sigma(\mathbf{x}_t \cdot \bm{\theta}_0 + \mathbf{x}_t \cdot \bm{\theta}(a_t) + b_1\kappa_{i,t}(a_t)+ b_2\kappa_{i,t}(a_t)^2).
\end{eqnarray}
The equation \ref{eq:predictor with fatigue} departs from eq. (\ref{eq:basic}) in two ways. First, the new predictor uses the metric for the level of advertising fatigue $\kappa_{i,t}(a_t)$. $\kappa_{i,t}(a_t)$ measures the estimated ad fatigue when a user $i$ is exposed to ad creative $a_t$ at time $t$. The specific calculation of $\kappa_{i,t}(a_t)$ is detailed below. Second, the new predictor take into account user $i$ while eq. (\ref{eq:basic}) does not contain subscript $i$. In the sense, the fatigue-aware predictor can personalize the ad creative selection.

$\kappa$ is included in the function as a quadratic form. However, both $b_1$ and $b_2$ can take any values.
While the marketing literature find inverse-U shape relationship described in Figure \ref{fig:two factor}, we \textbf{do not} presume the relationship between the two. The actual values of weights $(b_1, b_2)$ for fatigue metric are learned from the data. With the new predictor the DSP server chooses ad creative $a_t$ from a set of candidates $\mathcal{A}_t$. The whole procedure is Algorithm \ref{alg:fatigue}.

\begin{algorithm}[htb]
\caption{Fatigue-Aware Ad Creative Selection (FA)}         
\label{alg:fatigue}                  
\begin{algorithmic}                  
\REQUIRE: $\mathcal{N}\left(\bm{\mu}(a), \alpha\bm{\Sigma}(a)\right)$, $\mathbf{h}_{i,t}$, $\bm{s}$
\FOR{$t = 1, \cdots, T$}
\STATE Take feature vector $\mathbf{x}_{t}$ and history for $i$ at $t$ $\mathbf{h}_{i, t}$
\FOR{$a_t = 1, \cdots, |\mathcal{A}_t|$}
\IF{Action specific distribution is available}
\STATE get a vector of similarity scores for $a_t$, i.e. $\mathbf{s}(a_t)$
\STATE Calculate $\kappa_{i, t}(a_t) = \mathbf{h}_{i,t} \cdot \mathbf{s}(a_t)$
\STATE Sample $\bm{\theta}(a_t) \sim \mathcal{N}(\bm{\mu}(a_t), \alpha\bm{\Sigma}(a_t))$
\STATE Calculate $\hat{r}_{t}(a_t)$ according to eq.(\ref{eq:predictor with fatigue}).
\ELSIF{Action-specific distribution is NOT available}
\STATE Same as Algorithm \ref{alg:contextual}
\ENDIF
\ENDFOR
\STATE Select $a^\ast = \argmax\limits_{a \in \mathcal{A}_t} \hat{r}_{t}(a)$
\STATE Observe $r_t \in \{0, 1\}$
\IF{$t$ is the last impression of the day}
\STATE Update $\mathcal{N}(\bm{\mu}(a), \alpha\bm{\Sigma}(a)), b_1, b_2$ on data from previous day
\ENDIF
\ENDFOR
\end{algorithmic}
\end{algorithm}

\subsection{The Calculation of Fatigue}
\label{subsec:calc fatigue}
\begin{figure}[htb]
  \centering
  \includegraphics[width=0.3\linewidth]{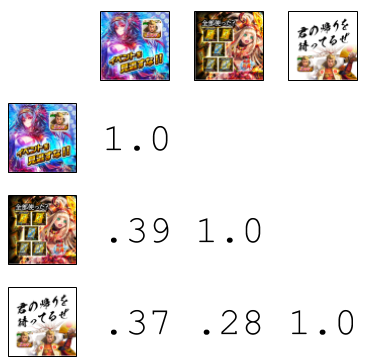}
  \caption{An example of similarity matrix.}
  \label{fig:similarity_matrix}
  \Description{}
\end{figure}

\begin{figure}[htb]
  \centering
  \includegraphics[width=0.9\linewidth]{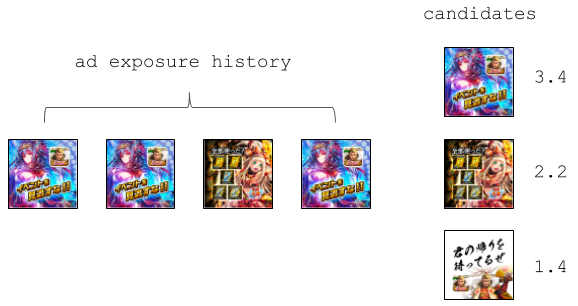}
  \caption{Calculation of fatigue. The values of fatigue depends on the history of ad exposure and chosen ad creative. In this case, blue ad creative has been seen three times and incur the highest fatigue.}
  \label{fig:fatigue_image}
  \Description{}
\end{figure}

Here, we detail the calculation of fatigue metric $\kappa$. The basic idea is that the level of fatigue is determined by the number of past exposure to the same or similar ad creatives. Suppose the similarities among three ad creatives are as shown in Figure \ref{fig:similarity_matrix}. 
Then the level of fatigue for each ad creative candidates can be sum of similarity between each candidate and creatives to which the user are exposed in the past. Figure \ref{fig:fatigue_image} shows an example. This user has seen the blue ad creative three times and the yellow creative one time. Then the ad fatigue becomes approximately 3.4 (3 for the same creative and .39 for yellow creative) when he sees the blue.

Formally, let $\mathbf{h}_{i, t}$ be a vector representation of history of ad exposures to user $i$ at time $t$. Each element in $\mathbf{h}_{i, t}$ represents the number of exposures to each creative. In the above example, $\mathbf{h_{i, t}} = (3, 1, 0)^T$. We record the histories for each user-advertiser pair. Hence, each user has multiple history vectors. The expected level of fatigue to ad creative $a_t$ is calculated as a weighted sum of history: $ \kappa_{i, t}(a_t) = \mathbf{h}_{i, t} \cdot \mathbf{s}(a_t)$, where the weight vector $\mathbf{s}(a_t)$ measure how similar $a_t$ and the other creatives are. That is, 
\begin{eqnarray}
\mathbf{s}(a_t) = (s(a_t, 1), s(a_t, 2), \cdots, s(a_t, |\mathcal{A}_t|))^T, s(j,k) \in [0,1].
\end{eqnarray}

\subsection{Calculation of Similarity}
\label{sec;calculation of similarity}
Each similarity score $s(a, a')$ is calculated as a weighted average of {\it text similarity} and {\it image similarity}. The text similarity is calculated by the cosine similarity of the bag of words (BoW) representations of the description texts.
We extracted the words from the texts by using Mecab\cite{Mecab}, which is the most commonly used morphological analyzer for the Japanese language, with NEologd dictionary\cite{neologd}, which is a frequently updated Japanese dictionary containing up-to-date words.
We used the gensim library\cite{gensim} for the BoW calculation. 

The image similarity is also calculated by the cosine similarity of the vector representation of the images.
The vector representation is extracted from each image by using pre-trained MobileNetV2\cite{MobileNetV2} implemented in Keras\cite{Keras}.
Here, we use the output from the last pooling layer as the representation.
We put three times higher weight to the text similarity because the text similarity (i.e., BoW representation) is much easier to interpret than the image similarity (i.e., weight of the neural network).

In a real world online display advertising environment, complicated models such as DNN are hard to be implemented because DSPs are required extremely short response time not to deteriorate user experience. By calculating similarity in advance, we can compress information from image and text into scalar variables it drastically reduces computing cost.

\subsection{Implementation}
\begin{figure}[h]
  \centering
  \includegraphics[width=\linewidth]{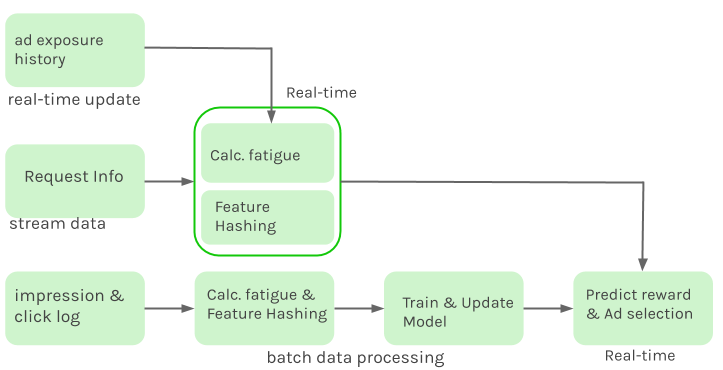}
  \caption{Summary of fatigue-aware ad creative selection system.}
  \label{fig:architecture}
  \Description{}
\end{figure}

The online advertising is an computationally complex environment. To deliver thousands of ad creatives to numerous users within unnoticeable duration, the system can spare only a few milliseconds to calculate fatigues for all the candidates of each impression. To save time, we first calculate the similarity scores for each pair of creatives off-line and store in the server. 
We update the histories for each user-advertiser pair in \textbf{real-time}. 
Owing to the extremely high frequency of impressions, recording all the ad creative exposures for all the users could lead to an explosion of the data volume, which needs to be capped at some point. 
Therefore, the system records the first impression of a user to each ad creative every minute. That is, if a user watches the same ad creative more than once within a minute, it is counted as one impression.\footnote{This data reduction affect less than 40\% of the users.} To further reduce the data volume and consider forgetting, the histories more than 24 hours before are deleted from the database. The whole system is described in Figure \ref{fig:architecture}. 

\section{Online Experiments}
\label{sec:Evaluation}
Since the reward is changed by the past intervention by the system, offline evaluation is not very meaningful in this case. Notice that replay method needs the exact same intervention history to each user in this case. Hence we run an online experiment in a production environment to examine the effectiveness of the proposed algorithm. In this section, we first describe the setting of the experiment and then show the results.

\subsection{Setting}
We deploy our Fatigue-Aware (FA) selection algorithm in a DSP provided by CyberAgent, inc., a Japan-based major online advertising company. Three advertising campaigns for three different mobile game titles are selected for the experiment according to various conditions \footnote{We used four campaigns but it turns out that one of them has only one creative.}. A part of impressions for these campaigns are used for the experiment. The sample size of the training data reaches a maximum of one million a day with the hundreds of thousand of features. To deal with the size and the sparsity of the training data, we employ negative down-sampling and the hashing trick\cite{weinberger_feature_2009}. We set sampling rate for negative data to 5\% and limit the length of features to $2^{24}$ for the system without fatigue parameters (Baseline) and $2^{24} + 2$ for FA by hashing-trick. The scalar parameter $\alpha$ that governs the degree of exploration of the algorithm are set to $0.01$ for both FA and Baseline. With the small value of $\alpha$, the Thompson sampling becomes {\it optimistic}\cite{chapelle_empirical_2011}. We apply stochastic gradient decent (SGDCliassifier in skleran package of Python) to obtain $\bm{\mu}$ with regularization parameter $\lambda = 0.0011$. The parameter vector is updated daily basis using a batch of data from the previous day. 
These hyper-parameters are tuned using the replay method \cite{li_contextual-bandit_2010} for Baseline. FA shares all the above setting except for the inclusion of the fatigue parameter. To deal with cold start problem, FA is trained on data from the existing algorithm in the pre-experiment period. During the experiment, it learns only data that it produces.

Baseline (Alg.\ref{alg:contextual}) and FA (Alg.\ref{alg:fatigue}), are tested against the result of randomly-chosen ad creative (Random) in a standard A/B testing. That is, they share the impressions equally based on the users' id and update their parameters based on the logs they generate. The experiment ran for one week. 

\subsection{Main Results}
Table \ref{tab:Main} shows the overall results of the experiments. Each metric is normalized with respect to the result of random algorithm (e.g., CTRs for FA and Baseline are divided by the CTR for Random). FA out-performed both baseline and random algorithm in CTR (click-through rate), CVR (conversion rate, the share of post-click conversions in all impressions), and Post-impression CVR including non-click conversions. The results clearly show that the proposed algorithm successfully increased both clicks and conversions. Baseline algorithm outperforms FA for post-click CVR. That is, while baseline collects less clicks than FA, baseline gets conversions more likely than FA once it gets click. But the difference is not statistically significant.

\begin{table}[htb]
\centering
\begin{tabular}{lrrrrr}
  \hline
 Alg. & Impressions & CTR & CVR & \shortstack{Post-Click \\CVR} & \shortstack{Post-Imp\\ CVR} \\ 
  \hline
  FA & 1,097,261 & \textbf{1.08}$^{\ast}$ & \textbf{1.12} & 1.04 & \textbf{1.09}$^{\ast\ast\ast}$ \\ 
  Baseline & 1,081,393 & 1.04 & 1.10 & \textbf{1.05} & 1.04 \\ 
  Rand & 1,087,931 & 1.00 & 1.00 & 1.00 & 1.00 \\ 
   \hline
\end{tabular}
\caption{The results of the experiments. The numbers are normalized except for the number of impressions. Normalization is performed with respect to the random algorithm. The results are for the aggregate of all three campaigns. Superscripts implies FA is statistically better than Baseline. $P<0.1:^\ast,  P <0.05:^{\ast\ast},  P<0.01 :^{\ast\ast\ast}$}
\label{tab:Main}
\end{table}

\subsection{Heterogeneity in campaigns}
To further examine the results, Table \ref{tab:campaigns} shows the CTRs for each campaign. The detailed results reveal that the baseline is not always superior to random. As for conversion rates, the performance of the baseline is poor in campaign A. On the other hand, our fatigue-aware algorithm stably outperforms the other two. The instability of baseline possibly comes from the data volume as campaign A has least size of data. With combination of small size of training data and sparsity, the estimates could be very unstable. In real-world production environment, the data volume cannot be controlled by experimenters. Still, fatigue bandit algorithm successfully learned the data.

\begin{table}[ht]
\centering
\begin{tabular}{lrrrr}
  \hline
& \multicolumn{3}{c}{Campaigns}\\
Alg & A & B & C \\ 
  \hline
FA & \textbf{1.21} & \textbf{1.04} & \textbf{1.32} \\ 
Baseline & 0.95 & 1.03 & 1.11 \\ 
Rand & 1.00 & 1.00 & 1.00 \\ 
   \hline
\end{tabular}
\caption{CTRs for three campaigns. Numbers are again normalized with respect to random algorithm.}
\label{tab:campaigns}
\end{table}

\section{Discussion}
\label{sec:Discussion}
\begin{figure}[htb]
 \includegraphics[width=\linewidth]{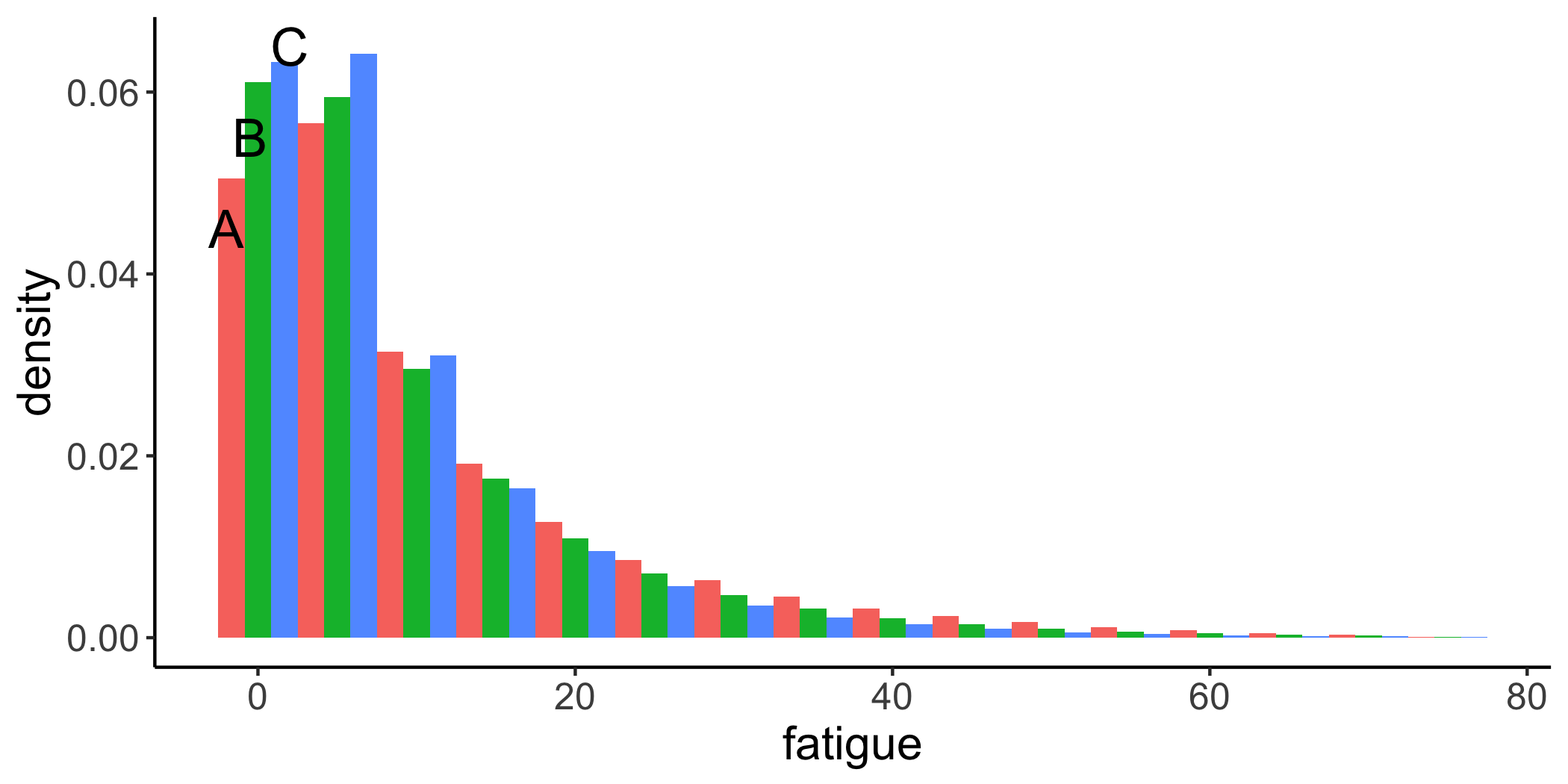}
  \caption{The distribution of fatigue level at each impression for three campaigns used in the experiment. The fatigue level is calculated according to Section \ref{subsec:calc fatigue}. Bin width is 5.}
  \label{fig:fatigue_dist}
  \Description{}
\end{figure}

In this section, we further investigate the results focusing on the fatigue. Figure \ref{fig:fatigue_dist} shows the distribution of user fatigue at time that the algorithm chooses the ad creative. While the distributions of fatigue are similar among three campaigns, Campaign A has the mean highest similarity score (Table \ref{tab:simiarity}) and higher fatigue level. The distribution is dense around 0 to 10, which is equivalent to 0 to 10 times exposure to same ad creative within past 24 hours.

\begin{table}[ht]
\centering
\begin{tabular}{cccc}
  \hline
Campaign & \# creatives & mean & sd \\ 
  \hline
A & 21 & 0.48 & 0.19 \\ 
  B & 12 & 0.39 & 0.09 \\ 
  C & 5 & 0.34 & 0.05 \\ 
   \hline
\end{tabular}
\caption{Summary of similarity scores for each campaign}
\label{tab:simiarity}
\end{table}

\begin{figure}[htb]
  \centering
  \includegraphics[width=4.5cm]{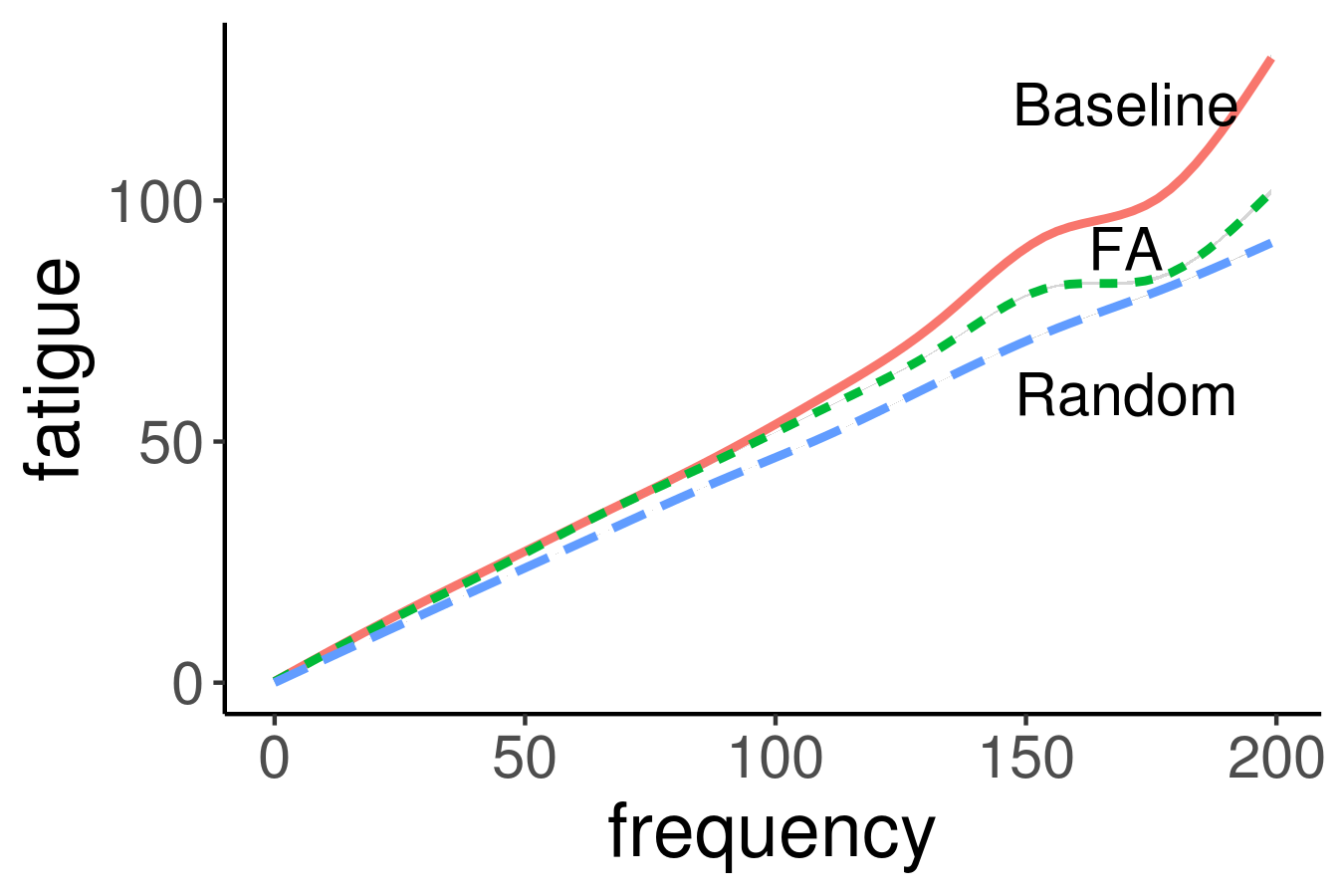}
  \caption{The relationship between the number of ad exposures in past 24 hours for each user (frequency) and the level of fatigue. Fitted values by local linear regressions of fatigue on frequency.}
  \label{fig:freq_fatigue}
  \Description{}
\end{figure}

Figure \ref{fig:freq_fatigue} shows the relation between the number of ad exposures and the the level of fatigue. Apparently, users feel least fatigue with random algorithm (dashed line). Random algorithm just distribute ad creatives evenly regardless of their performance while two other algorithms choose ``better'' ad creative to maximize clicks. We see the clear difference between fatigue-aware (dotted line) and baseline (solid line). Fatigue-aware algorithm tries to save the accumulation of fatigue by changing ad creatives when the level of fatigue is high. The fatigue-aware algorithm indeed works as we expected.

\begin{figure}[htb]
  \centering
  \begin{minipage}{.4\linewidth}
  \includegraphics[width=\linewidth]{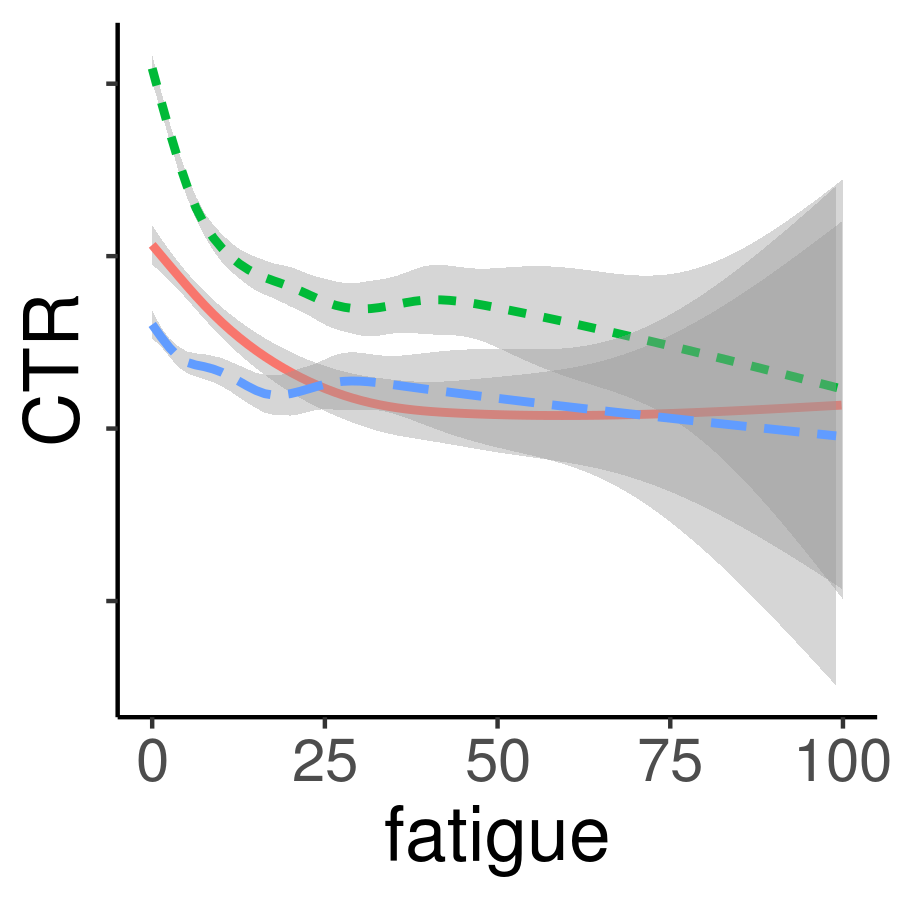}
    \end{minipage}
\begin{minipage}{.4\linewidth}
 \includegraphics[width=\linewidth]{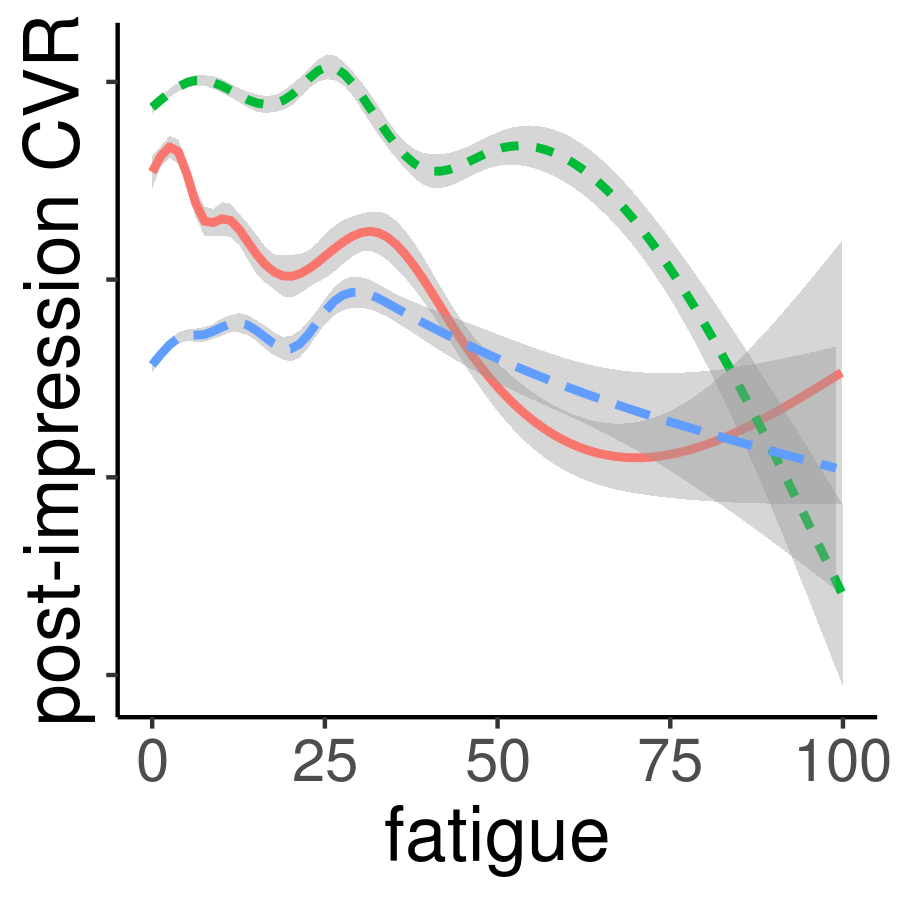}
    \end{minipage}
  \caption{The relationship between the level of fatigue and CTR (left) and post-impression CVR (right). Local linear regression of CTR/CVR on fatigue. Data is from random algorithm. Each line represents each campaign.}
  \label{fig:cv_fatigue}
  \Description{}
\end{figure}

Next, we check the association between the level of fatigue and the metrices (CTR and CVR). Figure \ref{fig:cv_fatigue} shows the local linear regressions CTR/CVR on fatigue. To eliminate bias generated by algorithms we use logs from random algorithm only. The two panels shows the difference between CTR and CVR. While CTR constantly decreases with the fatigue, CVR shows more complex relationship. In other words, we see only negative effects of fatigue on CTR but find both positive and negative effects for CVR. Possible explanation is that while audience tend to click on new and fresh ad creative for simple curiosity, they do not login game until they understand the message brought by ad creative after repetitive exposures. Hence, it is possible that the DSP should change the algorithm for different goals. This could explain why Baseline is better than FA in terms of post-click CVR. Baseline tends to show the same ad creatives consistently and gets more chance to get conversions once it gets clicks. But overall performance (CTR and CVR) is in favor of FA.

\section{Conclusion}
\label{sec:Conclusion}
In this paper, we proposed a fatigue-aware ad creative selection algorithm that explicitly considered the changes in users' attitude by the repetitive ad exposures. We introduced a numerical measure of the level of fatigue and set up easy-to-implement and efficient algorithm that changes ad creative according to the level of the fatigue. The proposed algorithm was applied to a running DSP in a production environment for one week. The results show the superiority of the proposed algorithm over the baseline algorithm.

\begin{acks}
The authors thank the Dynalyst team of CyberAgent, especially Hidetoshi Kawase, Yuta Kurosaki, and Shuhei Kimura for providing immense support to perform the experiments, and Kazuki Taniguchi, Masahiro Nomura, Kota Yamaguchi, Mayu Otani (AILab), Takanori Maehara (RIKEN), and Junpei Komiyama (NYU) for helpful advice.
\end{acks}

\newpage
\bibliographystyle{ACM-Reference-Format}
\bibliography{fatiguebandit}


\begin{thebibliography}{31}


\ifx \showCODEN    \undefined \def \showCODEN     #1{\unskip}     \fi
\ifx \showDOI      \undefined \def \showDOI       #1{#1}\fi
\ifx \showISBNx    \undefined \def \showISBNx     #1{\unskip}     \fi
\ifx \showISBNxiii \undefined \def \showISBNxiii  #1{\unskip}     \fi
\ifx \showISSN     \undefined \def \showISSN      #1{\unskip}     \fi
\ifx \showLCCN     \undefined \def \showLCCN      #1{\unskip}     \fi
\ifx \shownote     \undefined \def \shownote      #1{#1}          \fi
\ifx \showarticletitle \undefined \def \showarticletitle #1{#1}   \fi
\ifx \showURL      \undefined \def \showURL       {\relax}        \fi
\providecommand\bibfield[2]{#2}
\providecommand\bibinfo[2]{#2}
\providecommand\natexlab[1]{#1}
\providecommand\showeprint[2][]{arXiv:#2}

\bibitem[\protect\citeauthoryear{Abrams and Vee}{Abrams and Vee}{2007}]%
        {abramsPersonalizedAdDelivery2007a}
\bibfield{author}{\bibinfo{person}{Zo{\"e} Abrams} {and} \bibinfo{person}{Erik
  Vee}.} \bibinfo{year}{2007}\natexlab{}.
\newblock \showarticletitle{Personalized Ad Delivery When Ads Fatigue: {{An}}
  Approximation Algorithm}. In \bibinfo{booktitle}{\emph{International
  {{Workshop}} on {{Web}} and {{Internet Economics}}}}.
  \bibinfo{publisher}{{Springer}}, \bibinfo{pages}{535--540}.
\newblock


\bibitem[\protect\citeauthoryear{Agarwal, Chen, and Elango}{Agarwal
  et~al\mbox{.}}{2009}]%
        {agarwalSpatiotemporalModelsEstimating2009a}
\bibfield{author}{\bibinfo{person}{Deepak Agarwal}, \bibinfo{person}{Bee-Chung
  Chen}, {and} \bibinfo{person}{Pradheep Elango}.}
  \bibinfo{year}{2009}\natexlab{}.
\newblock \showarticletitle{Spatio-Temporal Models for Estimating Click-through
  Rate}. In \bibinfo{booktitle}{\emph{Proceedings of the 18th International
  Conference on {{World}} Wide Web - {{WWW}} '09}}. \bibinfo{publisher}{{ACM
  Press}}, \bibinfo{address}{{Madrid, Spain}}, \bibinfo{pages}{21}.
\newblock
\showISBNx{978-1-60558-487-4}
\urldef\tempurl%
\url{https://doi.org/10.1145/1526709.1526713}
\showDOI{\tempurl}


\bibitem[\protect\citeauthoryear{Bietti, Agarwal, and Langford}{Bietti
  et~al\mbox{.}}{2018}]%
        {bietti_contextual_2018}
\bibfield{author}{\bibinfo{person}{Alberto Bietti}, \bibinfo{person}{Alekh
  Agarwal}, {and} \bibinfo{person}{John Langford}.}
  \bibinfo{year}{2018}\natexlab{}.
\newblock \showarticletitle{A {Contextual} {Bandit} {Bake}-off}.
\newblock \bibinfo{journal}{\emph{arXiv:1802.04064 [cs, stat]}}
  (\bibinfo{date}{Feb.} \bibinfo{year}{2018}).
\newblock
\urldef\tempurl%
\url{http://arxiv.org/abs/1802.04064}
\showURL{%
\tempurl}
\newblock
\shownote{arXiv: 1802.04064.}


\bibitem[\protect\citeauthoryear{Bishop}{Bishop}{2006}]%
        {bishopPatternRecognitionMachine2006}
\bibfield{author}{\bibinfo{person}{Christopher~M. Bishop}.}
  \bibinfo{year}{2006}\natexlab{}.
\newblock \bibinfo{booktitle}{\emph{Pattern Recognition and Machine Learning}}.
\newblock \bibinfo{publisher}{{springer}}.
\newblock


\bibitem[\protect\citeauthoryear{Cao and Sun}{Cao and Sun}{2019}]%
        {cao_dynamic_2019}
\bibfield{author}{\bibinfo{person}{Junyu Cao} {and} \bibinfo{person}{Wei Sun}.}
  \bibinfo{year}{2019}\natexlab{}.
\newblock \showarticletitle{Dynamic {{Learning}} of {{Sequential Choice Bandit
  Problem}} under {{Marketing Fatigue}}}.
\newblock \bibinfo{journal}{\emph{Proceedings of the AAAI Conference on
  Artificial Intelligence}}  \bibinfo{volume}{33} (\bibinfo{date}{July}
  \bibinfo{year}{2019}), \bibinfo{pages}{3264--3271}.
\newblock
\showISSN{2374-3468}
\urldef\tempurl%
\url{https://doi.org/10.1609/aaai.v33i01.33013264}
\showDOI{\tempurl}


\bibitem[\protect\citeauthoryear{Chakrabarti, Kumar, Radlinski, and
  Upfal}{Chakrabarti et~al\mbox{.}}{2009}]%
        {chakrabarti_mortal_nodate}
\bibfield{author}{\bibinfo{person}{Deepayan Chakrabarti}, \bibinfo{person}{Ravi
  Kumar}, \bibinfo{person}{Filip Radlinski}, {and} \bibinfo{person}{Eli
  Upfal}.} \bibinfo{year}{2009}\natexlab{}.
\newblock \showarticletitle{Mortal Multi-Armed Bandits}.
\newblock In \bibinfo{booktitle}{\emph{Advances in Neural Information
  Processing Systems 21}}, \bibfield{editor}{\bibinfo{person}{D.~Koller},
  \bibinfo{person}{D.~Schuurmans}, \bibinfo{person}{Y.~Bengio}, {and}
  \bibinfo{person}{L.~Bottou}} (Eds.). \bibinfo{publisher}{Curran Associates,
  Inc.}, \bibinfo{pages}{273--280}.
\newblock
\urldef\tempurl%
\url{http://papers.nips.cc/paper/3580-mortal-multi-armed-bandits.pdf}
\showURL{%
\tempurl}


\bibitem[\protect\citeauthoryear{Chapelle and Li}{Chapelle and Li}{2011}]%
        {chapelle_empirical_2011}
\bibfield{author}{\bibinfo{person}{Olivier Chapelle} {and}
  \bibinfo{person}{Lihong Li}.} \bibinfo{year}{2011}\natexlab{}.
\newblock \showarticletitle{An {Empirical} {Evaluation} of {Thompson}
  {Sampling}}. In \bibinfo{booktitle}{\emph{Advances in neural information
  processing systems}}. \bibinfo{address}{Granada, Spain},
  \bibinfo{pages}{2249--2257}.
\newblock


\bibitem[\protect\citeauthoryear{Chatterjee, Hoffman, and Novak}{Chatterjee
  et~al\mbox{.}}{2003}]%
        {chatterjeeModelingClickstreamImplications2003}
\bibfield{author}{\bibinfo{person}{Patrali Chatterjee},
  \bibinfo{person}{Donna~L. Hoffman}, {and} \bibinfo{person}{Thomas~P. Novak}.}
  \bibinfo{year}{2003}\natexlab{}.
\newblock \showarticletitle{Modeling the {{Clickstream}}: {{Implications}} for
  {{Web}}-{{Based Advertising Efforts}}}.
\newblock \bibinfo{journal}{\emph{Marketing Science}} \bibinfo{volume}{22},
  \bibinfo{number}{4} (\bibinfo{date}{Nov.} \bibinfo{year}{2003}),
  \bibinfo{pages}{520--541}.
\newblock
\showISSN{0732-2399, 1526-548X}
\urldef\tempurl%
\url{https://doi.org/10.1287/mksc.22.4.520.24906}
\showDOI{\tempurl}


\bibitem[\protect\citeauthoryear{Chen, Yang, and Smith}{Chen
  et~al\mbox{.}}{2016}]%
        {chen_effects_2016}
\bibfield{author}{\bibinfo{person}{Jiemiao Chen}, \bibinfo{person}{Xiaojing
  Yang}, {and} \bibinfo{person}{Robert~E. Smith}.}
  \bibinfo{year}{2016}\natexlab{}.
\newblock \showarticletitle{The effects of creativity on advertising wear-in
  and wear-out}.
\newblock \bibinfo{journal}{\emph{Journal of the Academy of Marketing Science}}
  \bibinfo{volume}{44}, \bibinfo{number}{3} (\bibinfo{date}{May}
  \bibinfo{year}{2016}), \bibinfo{pages}{334--349}.
\newblock
\showISSN{0092-0703, 1552-7824}
\urldef\tempurl%
\url{https://doi.org/10.1007/s11747-014-0414-5}
\showDOI{\tempurl}


\bibitem[\protect\citeauthoryear{Chollet and others.}{Chollet and
  others.}{2015}]%
        {Keras}
\bibfield{author}{\bibinfo{person}{Fran\c{c}ois Chollet} {and}
  \bibinfo{person}{others.}} \bibinfo{year}{2015}\natexlab{}.
\newblock \showarticletitle{Keras}.
\newblock  (\bibinfo{year}{2015}).
\newblock
\urldef\tempurl%
\url{https://keras.io}
\showURL{%
\tempurl}


\bibitem[\protect\citeauthoryear{Gupta, Das, Pandey, and Narayanan}{Gupta
  et~al\mbox{.}}{2012}]%
        {guptaFactoringExposureDisplay2012}
\bibfield{author}{\bibinfo{person}{Neha Gupta}, \bibinfo{person}{Abhimanyu
  Das}, \bibinfo{person}{Sandeep Pandey}, {and} \bibinfo{person}{Vijay~K.
  Narayanan}.} \bibinfo{year}{2012}\natexlab{}.
\newblock \showarticletitle{Factoring {{Past Exposure}} in {{Display
  Advertising Targeting}}}. In \bibinfo{booktitle}{\emph{Proceedings of the
  18th {{ACM SIGKDD International Conference}} on {{Knowledge Discovery}} and
  {{Data Mining}}}} \emph{(\bibinfo{series}{{{KDD}} '12})}.
  \bibinfo{publisher}{{ACM}}, \bibinfo{address}{{New York, NY, USA}},
  \bibinfo{pages}{1204--1212}.
\newblock
\showISBNx{978-1-4503-1462-6}
\urldef\tempurl%
\url{https://doi.org/10.1145/2339530.2339719}
\showDOI{\tempurl}


\bibitem[\protect\citeauthoryear{Komiyama and Qin}{Komiyama and Qin}{2014}]%
        {liu_time-decaying_2014}
\bibfield{author}{\bibinfo{person}{Junpei Komiyama} {and} \bibinfo{person}{Tao
  Qin}.} \bibinfo{year}{2014}\natexlab{}.
\newblock \showarticletitle{Time-{Decaying} {Bandits} for {Non}-stationary
  {Systems}}.
\newblock In \bibinfo{booktitle}{\emph{Web and {Internet} {Economics}}},
  \bibfield{editor}{\bibinfo{person}{Tie-Yan Liu}, \bibinfo{person}{Qi~Qi},
  {and} \bibinfo{person}{Yinyu Ye}} (Eds.). Vol.~\bibinfo{volume}{8877}.
  \bibinfo{publisher}{Springer International Publishing},
  \bibinfo{address}{Cham}, \bibinfo{pages}{460--466}.
\newblock
\showISBNx{978-3-319-13128-3 978-3-319-13129-0}
\urldef\tempurl%
\url{https://doi.org/10.1007/978-3-319-13129-0_40}
\showDOI{\tempurl}


\bibitem[\protect\citeauthoryear{Lattimore and Szepesv´ari}{Lattimore and
  Szepesv´ari}{2019}]%
        {lattimore_bandit_2019}
\bibfield{author}{\bibinfo{person}{Tor Lattimore} {and} \bibinfo{person}{Csaba
  Szepesv´ari}.} \bibinfo{year}{2019}\natexlab{}.
\newblock \bibinfo{title}{Bandit {Algorithms}}.
\newblock
\newblock
\urldef\tempurl%
\url{https://tor-lattimore.com/downloads/book/book.pdf}
\showURL{%
\tempurl}


\bibitem[\protect\citeauthoryear{Lee, Lakshmanan, Tiwari, and Shah}{Lee
  et~al\mbox{.}}{2014}]%
        {leeModelingImpressionDiscounting2014}
\bibfield{author}{\bibinfo{person}{Pei Lee}, \bibinfo{person}{Laks~V.S.
  Lakshmanan}, \bibinfo{person}{Mitul Tiwari}, {and} \bibinfo{person}{Sam
  Shah}.} \bibinfo{year}{2014}\natexlab{}.
\newblock \showarticletitle{Modeling Impression Discounting in Large-Scale
  Recommender Systems}. In \bibinfo{booktitle}{\emph{Proceedings of the 20th
  {{ACM SIGKDD}} International Conference on {{Knowledge}} Discovery and Data
  Mining - {{KDD}} '14}}. \bibinfo{publisher}{{ACM Press}},
  \bibinfo{address}{{New York, New York, USA}}, \bibinfo{pages}{1837--1846}.
\newblock
\showISBNx{978-1-4503-2956-9}
\urldef\tempurl%
\url{https://doi.org/10.1145/2623330.2623356}
\showDOI{\tempurl}


\bibitem[\protect\citeauthoryear{Lei, Tewari, and Murphy}{Lei
  et~al\mbox{.}}{2017}]%
        {lei_actor-critic_2017}
\bibfield{author}{\bibinfo{person}{Huitian Lei}, \bibinfo{person}{Ambuj
  Tewari}, {and} \bibinfo{person}{Susan~A. Murphy}.}
  \bibinfo{year}{2017}\natexlab{}.
\newblock \showarticletitle{An {Actor}-{Critic} {Contextual} {Bandit}
  {Algorithm} for {Personalized} {Mobile} {Health} {Interventions}}.
\newblock \bibinfo{journal}{\emph{arXiv:1706.09090 [cs, stat]}}
  (\bibinfo{date}{June} \bibinfo{year}{2017}).
\newblock
\urldef\tempurl%
\url{http://arxiv.org/abs/1706.09090}
\showURL{%
\tempurl}
\newblock
\shownote{arXiv: 1706.09090.}


\bibitem[\protect\citeauthoryear{Levine, Koby, and Monner}{Levine
  et~al\mbox{.}}{2017}]%
        {levine_rotting_2017}
\bibfield{author}{\bibinfo{person}{Nir Levine}, \bibinfo{person}{Clammer Koby},
  {and} \bibinfo{person}{Shie Monner}.} \bibinfo{year}{2017}\natexlab{}.
\newblock \showarticletitle{Rotting {Bandits}}.
\newblock \bibinfo{journal}{\emph{Advances in Neural Information Processing
  Systems}}  \bibinfo{volume}{30} (\bibinfo{year}{2017}).
\newblock
\urldef\tempurl%
\url{http://papers.nips.cc/paper/6900-rotting-bandits}
\showURL{%
\tempurl}


\bibitem[\protect\citeauthoryear{Lewis}{Lewis}{2015}]%
        {lewis_worn-out_2015}
\bibfield{author}{\bibinfo{person}{Randall~A Lewis}.}
  \bibinfo{year}{2015}\natexlab{}.
\newblock \showarticletitle{Worn-{Out} or {Just} {Getting} {Started}? {The}
  {Impact} of {Frequency} in {Online} {Display} {Advertising}}.
  \bibinfo{address}{Boston, Massachusetts, USA}.
\newblock


\bibitem[\protect\citeauthoryear{Li, Chu, Langford, and Schapire}{Li
  et~al\mbox{.}}{2010}]%
        {li_contextual-bandit_2010}
\bibfield{author}{\bibinfo{person}{Lihong Li}, \bibinfo{person}{Wei Chu},
  \bibinfo{person}{John Langford}, {and} \bibinfo{person}{Robert~E. Schapire}.}
  \bibinfo{year}{2010}\natexlab{}.
\newblock \showarticletitle{A {Contextual}-{Bandit} {Approach} to
  {Personalized} {News} {Article} {Recommendation}}.
\newblock \bibinfo{journal}{\emph{Proceedings of the 19th international
  conference on World wide web - WWW '10}} (\bibinfo{year}{2010}),
  \bibinfo{pages}{661}.
\newblock
\urldef\tempurl%
\url{https://doi.org/10.1145/1772690.1772758}
\showDOI{\tempurl}
\newblock
\shownote{arXiv: 1003.0146.}


\bibitem[\protect\citeauthoryear{Ma, Liu, and Shen}{Ma et~al\mbox{.}}{2016}]%
        {maUserFatigueOnline2016a}
\bibfield{author}{\bibinfo{person}{Hao Ma}, \bibinfo{person}{Xueqing Liu},
  {and} \bibinfo{person}{Zhihong Shen}.} \bibinfo{year}{2016}\natexlab{}.
\newblock \showarticletitle{User {{Fatigue}} in {{Online News
  Recommendation}}}. In \bibinfo{booktitle}{\emph{Proceedings of the 25th
  {{International Conference}} on {{World Wide Web}}}}
  \emph{(\bibinfo{series}{{{WWW}} '16})}. \bibinfo{publisher}{{International
  World Wide Web Conferences Steering Committee}}, \bibinfo{address}{{Republic
  and Canton of Geneva, Switzerland}}, \bibinfo{pages}{1363--1372}.
\newblock
\showISBNx{978-1-4503-4143-1}
\urldef\tempurl%
\url{https://doi.org/10.1145/2872427.2874813}
\showDOI{\tempurl}


\bibitem[\protect\citeauthoryear{Matsumoto}{Matsumoto}{[n. d.]}]%
        {Mecab}
\bibfield{author}{\bibinfo{person}{Kitauchi A. Yamashita T. Hirano Y. Matsuda
  H.-Takaoka K. Asahara~M. Matsumoto, Y.}} \bibinfo{year}{[n. d.]}\natexlab{}.
\newblock \showarticletitle{Japanese morphological analysis system ChaSen
  version 2.0 manual.}
\newblock  (\bibinfo{year}{[n. d.]}).
\newblock


\bibitem[\protect\citeauthoryear{Pechman and Stewart}{Pechman and
  Stewart}{1988}]%
        {pechmanAdvertisingRepetitionCritical1988}
\bibfield{author}{\bibinfo{person}{Corneilia Pechman} {and}
  \bibinfo{person}{David~W. Stewart}.} \bibinfo{year}{1988}\natexlab{}.
\newblock \showarticletitle{Advertising {{Repetition}}: {{A Critical Review}}
  of {{Wearin}} and {{Wearout}}.}
\newblock \bibinfo{journal}{\emph{Current issues and research in advertising}}
  \bibinfo{volume}{11}, \bibinfo{number}{1-2} (\bibinfo{year}{1988}),
  \bibinfo{pages}{285--329}.
\newblock


\bibitem[\protect\citeauthoryear{{\v R}eh{\r u}{\v r}ek and Sojka}{{\v R}eh{\r
  u}{\v r}ek and Sojka}{2010}]%
        {gensim}
\bibfield{author}{\bibinfo{person}{R. {\v R}eh{\r u}{\v r}ek} {and}
  \bibinfo{person}{P Sojka}.} \bibinfo{year}{2010}\natexlab{}.
\newblock \showarticletitle{Software Framework for Topic Modelling with Large
  Corpora}. \bibinfo{publisher}{Proceedings of the LREC 2010 Workshop on New
  Challenges for NLP Frameworks.}
\newblock
\urldef\tempurl%
\url{http://is.muni.cz/publication/884893/en}
\showURL{%
\tempurl}


\bibitem[\protect\citeauthoryear{Sandler}{Sandler}{2018}]%
        {MobileNetV2}
\bibfield{author}{\bibinfo{person}{Howard A. Zhu M. Zhmoginov A. Chen L.-C
  Sandler, M.}} \bibinfo{year}{2018}\natexlab{}.
\newblock \showarticletitle{MobileNetV2: Inverted Residuals and Linear
  Bottlenecks.}
\newblock \bibinfo{journal}{\emph{arXiv:1801.04381 [cs]}}
  (\bibinfo{year}{2018}).
\newblock


\bibitem[\protect\citeauthoryear{Sato}{Sato}{2017}]%
        {neologd}
\bibfield{author}{\bibinfo{person}{Hashimoto T. Okumura~M Sato, T.}}
  \bibinfo{year}{2017}\natexlab{}.
\newblock \showarticletitle{Implementation of a word segmentation dictionary
  called mecab-ipadic-NEologd and study on how to use it effectively for
  information retrieval (in Japanese).}
\newblock \bibinfo{journal}{\emph{In Proceedings of the Twenty-three Annual
  Meeting of the Association for Natural Language Processing.}}
\newblock


\bibitem[\protect\citeauthoryear{Schmidt and Eisend}{Schmidt and
  Eisend}{2015}]%
        {schmidt_advertising_2015}
\bibfield{author}{\bibinfo{person}{Susanne Schmidt} {and}
  \bibinfo{person}{Martin Eisend}.} \bibinfo{year}{2015}\natexlab{}.
\newblock \showarticletitle{Advertising {Repetition}: {A} {Meta}-{Analysis} on
  {Effective} {Frequency} in {Advertising}}.
\newblock \bibinfo{journal}{\emph{Journal of Advertising}}
  \bibinfo{volume}{44}, \bibinfo{number}{4} (\bibinfo{date}{Oct.}
  \bibinfo{year}{2015}), \bibinfo{pages}{415--428}.
\newblock
\showISSN{0091-3367, 1557-7805}
\urldef\tempurl%
\url{https://doi.org/10.1080/00913367.2015.1018460}
\showDOI{\tempurl}


\bibitem[\protect\citeauthoryear{Schumann, Petty, and Scott~Clemons}{Schumann
  et~al\mbox{.}}{1990}]%
        {schumannPredictingEffectivenessDifferent1990}
\bibfield{author}{\bibinfo{person}{David~W. Schumann},
  \bibinfo{person}{Richard~E. Petty}, {and} \bibinfo{person}{D.
  Scott~Clemons}.} \bibinfo{year}{1990}\natexlab{}.
\newblock \showarticletitle{Predicting the Effectiveness of Different
  Strategies of Advertising Variation: {{A}} Test of the Repetition-Variation
  Hypotheses}.
\newblock \bibinfo{journal}{\emph{Journal of Consumer Research}}
  \bibinfo{volume}{17}, \bibinfo{number}{2} (\bibinfo{year}{1990}),
  \bibinfo{pages}{192--202}.
\newblock


\bibitem[\protect\citeauthoryear{Tang, Jiang, Li, and Li}{Tang
  et~al\mbox{.}}{2014}]%
        {tangEnsembleContextualBandits2014}
\bibfield{author}{\bibinfo{person}{Liang Tang}, \bibinfo{person}{Yexi Jiang},
  \bibinfo{person}{Lei Li}, {and} \bibinfo{person}{Tao Li}.}
  \bibinfo{year}{2014}\natexlab{}.
\newblock \showarticletitle{Ensemble Contextual Bandits for Personalized
  Recommendation}. In \bibinfo{booktitle}{\emph{Proceedings of the 8th {{ACM
  Conference}} on {{Recommender}} Systems - {{RecSys}} '14}}.
  \bibinfo{publisher}{{ACM Press}}, \bibinfo{address}{{Foster City, Silicon
  Valley, California, USA}}, \bibinfo{pages}{73--80}.
\newblock
\showISBNx{978-1-4503-2668-1}
\urldef\tempurl%
\url{https://doi.org/10.1145/2645710.2645732}
\showDOI{\tempurl}


\bibitem[\protect\citeauthoryear{Tang, Rosales, Singh, and Agarwal}{Tang
  et~al\mbox{.}}{2013}]%
        {tang_automatic_2013}
\bibfield{author}{\bibinfo{person}{Liang Tang}, \bibinfo{person}{Romer
  Rosales}, \bibinfo{person}{Ajit Singh}, {and} \bibinfo{person}{Deepak
  Agarwal}.} \bibinfo{year}{2013}\natexlab{}.
\newblock \showarticletitle{Automatic ad format selection via contextual
  bandits}. In \bibinfo{booktitle}{\emph{Proceedings of the 22nd {ACM}
  international conference on {Conference} on information \& knowledge
  management - {CIKM} '13}}. \bibinfo{publisher}{ACM Press},
  \bibinfo{address}{San Francisco, California, USA},
  \bibinfo{pages}{1587--1594}.
\newblock
\showISBNx{978-1-4503-2263-8}
\urldef\tempurl%
\url{https://doi.org/10.1145/2505515.2514700}
\showDOI{\tempurl}


\bibitem[\protect\citeauthoryear{Unnava and Burnkrant}{Unnava and
  Burnkrant}{1991}]%
        {unnavaEffectsRepeatingVaried1991a}
\bibfield{author}{\bibinfo{person}{H.~Rao Unnava} {and}
  \bibinfo{person}{Robert~E. Burnkrant}.} \bibinfo{year}{1991}\natexlab{}.
\newblock \showarticletitle{Effects of {{Repeating Varied Ad Executions}} on
  {{Brand Name Memory}}}.
\newblock \bibinfo{journal}{\emph{Journal of Marketing Research}}
  \bibinfo{volume}{28}, \bibinfo{number}{4} (\bibinfo{date}{Nov.}
  \bibinfo{year}{1991}), \bibinfo{pages}{406--416}.
\newblock
\showISSN{0022-2437, 1547-7193}
\urldef\tempurl%
\url{https://doi.org/10.1177/002224379102800403}
\showDOI{\tempurl}


\bibitem[\protect\citeauthoryear{Weinberger, Dasgupta, Langford, Smola, and
  Attenberg}{Weinberger et~al\mbox{.}}{2009}]%
        {weinberger_feature_2009}
\bibfield{author}{\bibinfo{person}{Kilian Weinberger}, \bibinfo{person}{Anirban
  Dasgupta}, \bibinfo{person}{John Langford}, \bibinfo{person}{Alex Smola},
  {and} \bibinfo{person}{Josh Attenberg}.} \bibinfo{year}{2009}\natexlab{}.
\newblock \showarticletitle{Feature hashing for large scale multitask
  learning}. In \bibinfo{booktitle}{\emph{Proceedings of the 26th {Annual}
  {International} {Conference} on {Machine} {Learning} - {ICML} '09}}.
  \bibinfo{publisher}{ACM Press}, \bibinfo{address}{Montreal, Quebec, Canada},
  \bibinfo{pages}{1--8}.
\newblock
\showISBNx{978-1-60558-516-1}
\urldef\tempurl%
\url{https://doi.org/10.1145/1553374.1553516}
\showDOI{\tempurl}


\bibitem[\protect\citeauthoryear{Zajonc}{Zajonc}{1968}]%
        {zajonc_attitudinal_1968}
\bibfield{author}{\bibinfo{person}{Robert~B. Zajonc}.}
  \bibinfo{year}{1968}\natexlab{}.
\newblock \showarticletitle{Attitudinal effects of mere exposure.}
\newblock \bibinfo{journal}{\emph{Journal of Personality and Social
  Psychology}} \bibinfo{volume}{9}, \bibinfo{number}{2, Pt.2}
  (\bibinfo{year}{1968}), \bibinfo{pages}{1--27}.
\newblock
\showISSN{1939-1315, 0022-3514}
\urldef\tempurl%
\url{https://doi.org/10.1037/h0025848}
\showDOI{\tempurl}


\end{thebibliography}

\end{document}